\documentclass[manuscript]{aastex}

\begin{document}

\title{Space Velocities of Southern Globular Clusters. IV.
First Results for Inner-Galaxy Clusters}

\author{Dana I. Dinescu\altaffilmark{1,2}, Terrence M.
Girard\altaffilmark{1}, William F. van Altena\altaffilmark{1}, and
Carlos L\'{o}pez\altaffilmark{3}}

\altaffiltext{1}{Astronomy Department, Yale University, P.O. Box 208101,
New Haven, CT 06520-8101 (dana@astro.yale.edu, girard@astro.yale.edu,
vanalten@astro.yale.edu}
\altaffiltext{2}{Astronomical Institute of the Romanian Academy, Str.
Cutitul de Argint 5, RO-75212, Bucharest 28, Romania}
\altaffiltext{3}{Yale Southern Observatory, University of San Juan,
Argentina}

\begin{abstract}
We have measured the absolute proper motions 
of four low-latitude, inner-Galaxy globular clusters.
These clusters are: NGC 6266 (M62), NGC 6304, NGC 6316 and
NGC 6723. The proper motions are on the $Hipparcos$ system, as 
no background extragalactic objects are found in these high-extinction
regions. The proper-motion uncertainties range between 
0.3 and 0.6 mas yr$^{-1}$.

We discuss the kinematics of these clusters and of three
additional bulge clusters --- NGC 6522, NGC 6528 and NFC 6553 ---
whose  proper motions with respect to bulge stars had been determined 
previously.
We find that all of the clusters have velocities that confine them to 
the bulge region. Of the three metal poor clusters ([Fe/H] $< -1.0$),
NGC 6522, and NGC 6723 have kinematics consistent with halo membership.
The third cluster, NGC 6266 however, appears to belong to a 
rotationally-supported system.
Of the four metal rich clusters ([Fe/H] $\ge -1.0$), 
NGC 6304 and NGC 6553 also have kinematics consistent with membership to
a rotationally-supported system. NGC 6528 has 
kinematics, metallicity and mass that argue in favor of a genuine
Milky-Way bar cluster. NGC 6316's kinematics indicate membership to a
hotter system than the bar.

\end{abstract}


\section{Introduction}

Our Galaxy's inner regions are a challenge for
Galactic structure studies, as they consist of a superposition of
stellar populations of various spatial, physical and kinematical properties,
all of which peak in density in these regions. In order to understand the 
stellar populations of the inner Galaxy,
one desires a tracer population that has, at the least,  
reliable measurements of distances, radial velocities, proper motions and 
metallicities, and, ideally, detailed chemical abundances and ages.
Globular clusters fulfill these requirements well. 
Their only drawback is their 
small number, when compared to that of field stars. 

There is still the question of whether the clusters are representative 
of the Galaxy's field stellar population. Many recent studies point to the
fact that the globular-cluster population does resemble in many ways
the old stellar field component.
Based on radial velocities and metallicities it has been shown that
globulars can be divided into thick disk and halo components
(Zinn 1985, Armandroff 1989). Later, Minniti (1995) studied the
density distribution of  metal rich clusters in the bulge region
($r_{GC} \le 3$ kpc)  \footnote[1]{Here, 
$r_{GC}$ represents the Galactocentric radius in a spherical 
coordinate system, and $R_{GC}$ the Galactocentric 
distance in a cylindrical coordinate system.}, and he concluded
 that most of these 
clusters belong to the bulge system rather than the disk.
More recent distance, radial-velocity, and metallicity
measurements of clusters in the bulge region
together with more detailed analyses
(Barbuy, Bica, \& Ortolani 1998,
C\^{o}t\'{e} 1999, Heitsch \& Richtler 1999; hereafter
HR99, Burkert \& Smith 1997; hereafter BS97)
suggest that some clusters may form a bar system that resembles the 
Milky Way bar (see Gerhard 2002 for a review of observations and 
characteristics of the Galactic bar).

Regarding clusters outside the bulge region but within the
solar circle, Zinn (1996) shows that intermediate metal-poor clusters
($-1.8 < $ [Fe/H] $ < -0.8$) have significant rotation 
($V_{rot} \sim 60$ km s$^{-1}$) which is
however, lower than that of the thick disk. This analysis included 
radial-velocity data alone and the typical Frenk \& White (1980) solution,
that can determine a mean rotation of a pre-selected group of
clusters. 

Adding tangential velocities as a piece of 
crucial information, Cudworth \& Hanson (1993) and later, Dinescu,
Girard \& van Altena (1999; Paper III) have unambiguously shown that
there is a metal-weak thick disk component in the cluster population, 
much like that seen in the local field stellar population (see 
Beers {\it et al.} 2002 for the most recent study).
And, the inner-halo rotation detected by Zinn (1996),
was also found in the studies that included 
tangential velocities (Paper III). This rotation is
qualitatively consistent with the flattened shape of 
the inner halo as determined from field
stars (e.g., Hartwick 1987, Chiba \& Beers 2001, Siegel {\it et al.} 2002).
To summarize, the globular-cluster system of our Galaxy appears to 
resemble the old stellar field component.

We set out to measure tangential velocities (i. e., absolute proper motions)
for a sample of fifteen clusters (Dinescu, Girard, \& van 
Altena 2002) primarily located within the solar
circle, in order to more accurately assign
clusters to the components of the inner Galaxy, and to 
better understand these components as regions where
old and massive stellar systems were produced/captured.

In this paper we present the first results for four globular clusters
located in the bulge region: NGC 6266 (M 62), NGC 6304, NGC 6316, and
NGC 6723. We discuss the kinematics of our four globular clusters
and of three additional clusters (NGC 6522, NGC 6528, and NGC 6553)
located in the bulge region that have previous proper-motion measurements,
in light of the Galactic components seen in the bulge region.

The proper-motion measurements are described in Section 2,
the velocity results are presented in Section 3, and discussed in Section 4,
and a brief summary is presented in Section 5.

\section{Measurements}

This project is part of the continuation of the Southern Proper Motion
Program (SPM, Platais {\it et al.} 1998), and a low-latitude supplement 
to the previous set of fifteen high-latitude globular clusters
measured by our group (Dinescu {\it et al.} 1997, 1999, Paper I and II
respectively). SPM plates are taken
with the 50-cm double astrograph at Cesco Observatory in El Leoncito,
Argentina (plate scale = 55.1 $\arcsec$/mm). 
The first-epoch survey, covering the sky south of declination 
$-17 \arcdeg$, was completed in the early 1970's.
The second-epoch survey was made in the 1990's, but after completing
roughly one third of the intended coverage, Kodak discontinued
production of the necessary 103a emulsion plates. The remainder of the
second-epoch survey will be obtained with a 4Kx4K CCD camera which has
recently been installed on the astrograph. 

The four clusters presented here lie within SPM fields which have both 
first and second epoch photographic plates.
The results in this paper are based on these plates, and on the
UCAC1 catalog (Zacharias {\it et al.} 2000, hereafter Z00).
Usually, there are 2 plate pairs (blue: 103aO, and visual: 103aG + OG515
filter) per SPM field, which cover an area of $6\fdg3$x$6\fdg3$.
Each SPM plate contains two exposures: a two-hour exposure, which
reaches to about $V = 18$, and an offset two-minute exposure,
which allows a tie-in to bright reference stars. During both exposures, an
objective grating is used, which produces a series of diffraction images
on either side of the central zero-order image. 

The program clusters were selected based on the limitations imposed by
the SPM survey. Thus, they are located in the southern sky, and 
within distances of 11 kpc from the Sun. A few additional
clusters at larger distances are in our program list because they happen
to be on the same SPM field with one or two more nearby clusters.

Since detailed descriptions of the measurement process and 
reduction procedures are given in Papers I and II, we will
only outline here the most important aspects. The photographic plates 
are scanned  with the Yale PDS microdensitometer,
in an object-by-object mode, at a pixel size of 12.7 microns.
We prepare an input catalog that contains all Hipparcos and Tycho 
(ESA 1997) stars,
$\sim 3000$ faint field stars selected from the USNO-A2.0 catalog
(Monet {\it et al.} 1998)
in the magnitude range 15 to 17, $\sim 150$ stars from the GSC 1.1 catalog
(Lasker {\it et al.} 1990)
with magnitudes fainter than those of the Tycho stars ($V = 11 - 14$),
and cluster stars. 
Hipparcos stars provide the correction to absolute proper motion,
while Tycho and GSC stars assure a good range in the magnitude
of various diffraction orders in order to model magnitude-dependent
systematics. For all of these relatively bright stars, we measure
long and short-exposure images, and diffraction images.

The faint stars from the USNO-A2.0 catalog allow a well-determined plate model
(see Paper I), and for these stars we measure only the long-exposure,
central order image. The list of cluster stars is determined using
the software package Sextractor (Bertin \& Arnouts 1996) from a 
preliminary scan in the cluster region of the best quality plate.
Cluster stars are selected in a region of radius 2-3 times the half-mass
radius as taken from Harris (1996; hereafter H96). Typically, this radius is 3-4 arcmin, with
the very central 1-2  arcmin totally unusable because of crowding. 
For the cluster stars we also measure only the long-exposure, 
central order image.

The dominant systematic effect in these
proper-motion measurements  is a magnitude-dependent shift in the image 
position owed to the nonlinear response of the photographic emulsions and
asymmetric image profiles caused by guiding errors during long exposures, 
imperfections in the optical system, polar misalignment of the telescope,
etc. We call this bias ``magnitude equation'', and we caution that it
is present in virtually all photographic plate material.
For our plate material for instance, over a range of 6 magnitudes 
(typical globular cluster stars have 
$V \ge 15$, and Hipparcos stars have $V \sim 9$), magnitude equation can 
produce deviations of 10-15 mas yr$^{-1}$ (Girard {\it et al.} 1998).
Note that, at a distance of 6 kpc, 1 mas yr$^{-1}$ corresponds to 
28 km s$^{-1}$. We use the diffraction images in order to
correct for magnitude equation according to the prescription 
described in Girard {\it et al.} (1998). In short, in absence
of magnitude equation, each diffraction-order pair of images would be
symmetrically displaced relative to the central order. Magnitude equation
produces a deviation of the position of the central order image from the
position expected from the geometry of the diffraction images,
as the central-order image is 4 magnitudes brighter than the 
diffraction images. Thus,
by measuring a sufficient number of stars whose central and diffraction images
are separable, we are able to deduce the exact form of the
magnitude equation on each SPM plate, and correct for it (see 
details and tests of this procedure in Girard {\it et al.} 1998).

The magnitude equation correction is of
significant size, and therefore, at the bright end, there is a 
non-negligible uncertainty in it. Thus, in the correction to 
absolute proper motion, we will use only Hipparcos stars
fainter than $V = 8$, to avoid residual magnitude equation in our
proper motions. Also, the blue plates used in this study have a poorer
determination of the magnitude equation than that of the visual 
plates, because the ``blue'' images are less sharp than the 
``visual'' ones. This is due to the fact that the blue plates 
are taken with no filter, and atmospheric seeing is generally
poorer in the blue than in the visual passband.
On some fields, in the blue passband only,
 we had to use the second-order diffraction 
images, because the first-order images were blended with the central order.

We also obtain photographic photometry by calibrating the instrumental
magnitudes with Tycho magnitudes, and with already published
CCD and/or photoelectric photometry in the cluster region. This latter 
photometry ensures a good calibration at the faint end.
Our photographic magnitudes have uncertainties of 0.15-0.25 mags.
There were two overlapping SPM fields for clusters NGC 6266, NGC 6304,
and NGC 6316. For these two fields, we have used CCD photometry from
Brocato {\it et al.} (1996) (NGC 6266), Rosenberg {\it et al.} (2000) 
(NGC 6266, and NGC 6304), and Ortolani {\it et al.} (2000) (NGC 6304).
For the SPM field containing NGC 6723, we have used photoelectric 
photometry from Alvarado {\it et al.} (1994) and CCD photometry 
from Rosenberg {\it et al.} (2000).

\section{Results}

\subsection{Proper Motions}

Table 1 summarizes our plate-pair solutions. We have also used 
the UCAC1 positions in combination with our first-epoch plates,
because they provide a longer baseline 
 ($ \sim 30$ years; mean epoch $\sim 1999$) than  
the SPM plates alone.  However, UCAC1 is less deep 
($ R \le 16$) than the SPM plate material, and
therefore solutions including this catalog miss
a number of faint cluster stars.
The denomination of each plate solution in Table 1 
is as follows: the number represents the SPM field number;
B1, Y1, represent the blue, and visual first-epoch 
plate respectively, and similarly B2, Y2 refer to the second-epoch plates;
ucac represents the corresponding SPM area selected from the UCAC1 catalog.

From a full plate solution that uses a large number of reference stars
($\sim 3000$), with a denser ring of stars around
the cluster (Paper I and II), we obtain relative proper motions
for all of the stars/images. The difference between the Hipparcos
absolute proper motions and the relative proper motions obtained
in our solution, represents the correction to absolute
proper motion to be applied to the relative proper motion
of the cluster stars. We calculate this proper-motion
difference for each SPM grating-image order of the 
Hipparcos stars. We plot these
differences as a function of color, magnitude and position on
the plate, to inspect for residual systematics. 
For each image order, we calculate the average and the 
scatter of these differences using probability plots (see Paper I and II).
The final correction to absolute proper motion is a weighted mean of
the correction given by each image order.  The weights are given by the
internal scatter within each image order (see also Paper II for cluster M4).
In some cases we simply discard 
image orders that we know are erroneous (poor image quality, 
uncertain  magnitude-equation correction). The number of Hipparcos
stars in each image order ranges between 30 and 90.
In Figure 1 we show, as an example, the correction to absolute proper motion
for plate-pair solutions 506Y1Y2 and 506y1-ucac
(left and right panel respectively;
$\mu_{\alpha}^{*} = \mu_{\alpha}$ cos $\delta$). Each point
represents the average of a given image order and is labeled
as follows: 0 - long-exposure central order, 1 - long-exposure first
order, 2 - long-exposure second order, 6 - short-exposure central order, 
7 - short-exposure first order, and 8 - short-exposure second-order
(The image order 8 is not shown here because it is a very poor measurement). 
The error bars are uncertainties in the averages as given by the 
internal error estimate of each image order. The weighted mean of all image orders
are also shown in  the figure, with the dark symbols. 
Overall, the 506Y1-ucac measurement appears to be better than the 506Y1Y2
from the scatter and the formal errors in each image order.
However, the 506Y1-ucac measurements are not independent
(as the SPM ones are), because for each image order on the 
first-epoch SPM plate there is a corresponding image-order 
measurement on the second epoch, but only one UCAC1 position.

The uncertainty in the mean cluster motion is dominated by 
the image centering error from the PDS scans and a relatively small number of 
cluster stars. We plot the centering error as a function of 
magnitude, and select those stars that follow a known 
sequence as determined from stars in the uncrowded region of the plate; a
centering-error upper limit of 2 microns is also adopted.
Stars with proper-motion sizes larger than 40 mas yr$^{-1}$
are excluded. 
The average relative cluster proper motion is a centering-error weighted mean.
We have also used probability plots to estimate this quantity,
and we found that it is similar to the weighted mean within the
uncertainties. The field star contamination does not affect
our mean cluster proper motion for the following reasons:
1) the cluster area is a very small region, where 
10 - 15 \% of the stars measured in the cluster area may belong to the field
(this estimate is based on the stellar density in the ring of
reference stars that surrounds the cluster, see also Paper I), and 2) 
the field proper-motion dispersion is large ($ \sim 10 $ mas yr$^{-1}$)
in the magnitude range $V  = 14 - 18$ of the reference stars; thus, 
the potential of systematically affecting our estimate of the mean 
cluster motion is negligible.
In Figure 2 we show relative proper motions in the region of
NGC 6304 as a function of magnitude, for visual-plate solutions
in SPM fields 506 and 507. The stars shown are those trimmed in 
centering-error magnitude space.
The horizontal lines represent the centering-error weighted means.
Using cluster stars, whose proper-motion dispersion reflects 
the measurement error only,  and assuming that the positional uncertainty 
on the SPM plates is the same at both epoch, we can
estimate the positional uncertainty for UCAC1 stars in the
magnitude range $ 14 \le V \le 17$. From our Y plate-pair solutions that
have the best quality images on SPM fields 506 and 507, 
we obtain a positional uncertainty between 98 and 110 mas
for UCAC1 stars in NGC 6304. These values
are larger than the standard error of UCAC1 (Z00) at $R = 16$: 70 mas.
This is perhaps due to crowding/blending in the cluster region due to
the 100 $\arcsec$/mm plate scale of the US Naval Observatory Twin Astrograph.

The absolute proper motion is the sum of the relative cluster 
proper motion and the correction to absolute as given by the Hipparcos stars
(also called zero point).
Uncertainties in the relative cluster proper motion range between
0.33 and 1.44 mas yr$^{-1}$, while in the zero point they range between
0.22 and 0.87 mas yr$^{-1}$.
The dominant error in the absolute proper-motion
uncertainty is that given by the relative cluster motion.
In Table 2 we summarize our absolute proper motions for each cluster
and plate solution; we also specify the number of cluster stars used in each 
solution. In Figure 3 we plot the results from Table 2, on 
a single scale for comparison purposes.

We note that some data points in Fig. 3 are not strictly independent,
because of the use of the same plate in two plate-pair solutions
(Table 1). This has been taken into account when we 
estimate the final absolute proper motion and associated uncertainty.
We do so by determining relative mean positions of 
each cluster from the 
absolute proper-motion measures (see Table 2) at the three different epochs,
as follows.
We adopt an arbitrary nominal position for the UCAC1 epoch. For each passband,
we predict a position at the 1st-SPM epoch using the SPM-ucac proper motions.
Then, using the SPM-only solution, we predict the 2nd-SPM epoch positions. 
From the proper-motion uncertainty, and the number of
cluster stars used in each plate-pair solution we are able to 
obtain positional uncertainties for the mean cluster position
at a given epoch. Then, we determine a (position-uncertainty) weighted fit
of the mean cluster position as a function of time. The slope and
its formal error from the fit yield the final absolute proper motion
and  its corresponding uncertainty. 

For NGC 6266, there are five independent position data points (see Table 1).
We obtain 
$\mu_{\alpha}^{*} =  -3.50 \pm 0.37$ and $\mu_{\delta} = -0.82 \pm 0.37$
mas yr$^{-1}$ for NGC 6266.
For NGC 6304, using seven 
independent mean cluster positions (Table 1), we obtain:
$\mu_{\alpha}^{*} =  -2.59 \pm 0.29$ and $\mu_{\delta} = -1.56 \pm 0.29$
mas yr$^{-1}$. NGC 6316, being the most distant cluster among our
four clusters, has only two reliable
(proper-motion uncertainties smaller than 1.5 mas yr$^{-1}$)
measurements provided by the 
visual plate pairs (Table 1, 2). These two measurements are independent,
and our final proper motion is a weighted mean of the two values 
from Table 2. We obtain
$\mu_{\alpha}^{*} =  -2.42 \pm 0.63$ and $\mu_{\delta} = -1.71 \pm 0.56$
mas yr$^{-1}$ for NGC 6316.
For NGC 6723, we apply the same procedure as for NGC 6266, and 6304.
According to the plate-pair solutions (Table 1) there are five 
mean cluster positions, of which those at the second SPM epoch are not 
independent, because of the mixed 440Y1B2 plate solution. We scale up our
formal proper-motion uncertainties by the factor
 $\sqrt{(N_{pts;all} - N_{fit})/(N_{pts;indep.} - N_{fit})}$ = 1.22,
where $N_{fit} = 2$.
We thus obtain $\mu_{\alpha}^{*} =  -0.17 \pm 0.45$ and $\mu_{\delta} = -2.16 \pm 0.50$ mas yr$^{-1}$ for NGC 6723.
The final absolute proper motions are represented with filled symbols 
in Fig. 3.

\subsection{Velocities}

In addition to our four clusters, we have selected three 
clusters located in the bulge region
that have previous proper-motion measurements. For these, the correction to an
inertial reference frame is based upon the assumption that the reference
stars are bulge stars, all located at the same (known) distance, and
their velocity with respect to the Galactic 
rest frame is, on the mean, zero. These clusters are NGC 6522
(Terndrup {\it et al.} 1998),
NGC 6528 (Feltzing \& Johnson 2002),
and NGC 6553 (Zoccali et al. 2001). The latter two studies are based
on HST data.
The Terndrup {\it et al.} (1998) study has a detailed description of 
the assumptions and biases of this method. An important assumption
is that of only a small deviation from the Galactic Center
(GC) in latitude and, most importantly, in
longitude (see Discussion Section for NGC 6553).
Here, we will rederive their space velocities according
to the published proper motions.

In Table 3 we summarize the basic parameters for deriving space velocities.
Distances and heliocentric radial velocities are from H96.
The numbers in parentheses represent the uncertainties. We adopt a nominal
distance uncertainty of 10\% of the distance.
We use the Solar peculiar motion $(U_{\odot}, V_{\odot}, W_{\odot})$ = 
(-11.0, 14.0, 7.5) km s$^{-1}$
(Ratnatunga, Bahcall, \& Casertano 1989), 
where $U$ is positive outward from the GC, 
$V$ is positive toward Galactic rotation,
and $W$ is positive toward the north Galactic pole. The Sun is located at
(X, Y, Z) = (8.0, 0.0, 0.0) kpc, and  the rotation velocity 
of the local standard of rest
is taken to be $V_{LSR} = 220.0$ km s$^{-1}$. In what follows we will
use $V'= V + V_{LSR}$, and velocity components in a cylindrical 
coordinate system of $\Pi$ (positive outward from GC) and $\Theta$ 
(positive toward Galactic rotation). For clusters located at small 
GC distances (i.e., $R_{GC}$), 
the cylindrical coordinate system is no longer appropriate, and
$\Pi$ and $\Theta$ components can be misleading.
In these cases,
the Cartesian system in the Galactic rest frame ($U,V',W$) is more suitable.
We will come back to this issue when we discuss the kinematics of each cluster.

Table 4 lists the (X, Y, Z) coordinates, the velocity components, 
and the rotation velocity expected from a solid body that has 
an angular velocity equal to the pattern speed of the Milky Way bar
at the location of the cluster
($\Omega_{b} = 60$ km s$^{-1}$ kpc$^{-1}$, see the following Section).

\section{Discussion}

All of the clusters under discussion reside within $\sim 3$ kpc
of the GC (Tab. 4). This is a very complex region dominated
by a bar/bulge where a traditional kinematical   
distinction between the disk and the bar/bulge may not be as meaningful
as in more distant regions from the GC,
where  disk and halo populations for instance are easily separated.
Dynamical criteria allow 
the existence of a kinematical disk only outside the corotation radius
(see e. g.,  Pfenniger \& Friedli 1991).
N-body simulations show that bars form easily as a gravitational 
instability in a cold disk, and, in order to survive, they have to
rotate (Pfenniger 1999). Bars then, can be regarded as ``thickened'' 
elliptical disks. They preserve the ``memory'' of the original
rotating axisymmetric disk, and therefore one may be able to
kinematically  distinguish between the bar/bulge and a 
pressure-supported system such as the halo.
Presumably, a halo particle will remain a halo particle, while a 
disk particle will become
a bar/bulge particle inside the corotation radius.
In N-body simulations, there is evidence of some 
resonant halo stars trapped in the bar, in particular
at corotation (Athanassoula 2002); 
however, for our study that considers very small
numbers, we believe that it is unlikely to find a halo globular
cluster on a resonant orbit with the Galactic bar.

A separation between bar and bulge is more ambiguous given
the variety of processes that may contribute to bulge formation
such as bar growth, bar dissolution, accretion and/or merging with satellites.
 
In our Galaxy, evidence for a bar 
comes from observations of motions of atomic and molecular gas,
from near-infrared photometry (NIR), IRAS sources and clump stars counts,
and stellar kinematics (see e.g., Gerhard 1996 for a review).  
The characteristics of the Milky Way bar are summarized in 
Gerhard (2002). The bar has a semimajor axis 
a = $3.1 - 3.5$ kpc, an axis ratio of 10:3-4:3, and an orientation angle
(angle between the semimajor axis and the Sun-GC line)
of $\phi_{bar} = 15\arcdeg - 35\arcdeg $. The pattern speed is
$\Omega_b = 50 - 60$ km s$^{-1}$ kpc$^{-1}$, and the corresponding
corotation radius is $R_{CR} = 3.3 - 5 $ kpc.
The Galactic bar seems to be a fairly massive system
(M$_{bar} \sim 10^{10}$ M$_{\odot}$; Stanek et al. 1997, Weiner \&
Sellwood 1999),
of an age of up to $\sim 6 - 7$ Gyr (as determined from Carbon
stars, Cole \& Weinberg 2002 and references therein).
Given these considerations, can we speak of a globular-cluster
bar system? Were these clusters trapped into bar-like orbits from
an initially disk-like configuration, as dissipationless N-body
simulations show? Or, were these clusters produced within the bar,
if the bar formed from gas rather than from a stellar distribution?
BS97 argued that low-mass (i.e. integrated absolute magnitude
M$_V \ge -7.9$), metal-rich ([Fe/H] $ > -0.8$)
clusters show spatial distribution and kinematics --- as derived from 
radial velocities ---
consistent with those of the Milky Way bar 
(see also C\^{o}t\'{e} 1999 and references therein).
Given the large mass of the bar, and its upper age limit, it is quite
conceivable that low-mass, metal-rich clusters formed within the bar.
Dating the clusters suggested by BS97 as belonging to a bar-like 
configuration could be very instructive.

In light of the above, we proceed to discuss the results
for the seven clusters under study, without attempting 
a detailed orbit integration that we defer to a future paper.

In Table 5 we list the cluster metallicity [Fe/H], integrated absolute
magnitude M$_V$ (H96), total velocity, escape velocity from the bulge, 
and orbit inclination. The orbit inclination is
$\Psi = 90\arcdeg -$ sin$^{-1}$(L$_z$/L), where L$_z$ is the angular
momentum perpendicular to the Galactic disk, and L is the total
angular momentum, at the present time. In contrast to Paper III, this
is not an orbit-averaged quantity, but an instantaneous one. 
The bulge potential is the Hernquist spheroid
described in Johnston, Spergel \& Hernquist (1995; JSH95).
This potential is a simple description of the inner Galaxy, and it 
is used only to estimate the escape velocity and not to derive
orbits. As total velocities are smaller than escape velocities,
all of the clusters are confined to the bulge region.

In Figure 4 we show the locations and motions of the clusters
in the Galactic plane
(top panel) and perpendicular to the Galactic plane (bottom panel).
The open symbols represent the metal poor clusters, 
the filled symbols the metal rich ones (see following subsections).
The dashed line represents the bulge region; it has a radius of 2.7 kpc,
which corresponds to an angle of $20\arcdeg$ on the sky from the GC
(from COBE and IRAS maps, and $R_{\odot} = 8.0$ kpc).
The ellipse represents the bar. Its characteristics are
those determined by Bissantz  \& Gerhard (2002) from a model that
is constrained by the dereddened COBE/DIRBE L-band luminosity distribution and
by the distribution of red clump stars in some bulge fields.
The bar has a semimajor axis of
3.5 kpc, an axis ratio of 10:3:3, and an orientation angle
of $20\arcdeg$.

\subsection{Metal Poor ([Fe/H] $< -1.0$) Clusters}

NGC 6522 and NGC 6723 have locations and velocity
components that produce highly inclined orbits (Tab. 5).
Their kinematics suggest these clusters are members of
the halo system. Their low
metallicities are in agreement with this assignment 
(e. g., Armandroff 1989).
We note that, within the uncertainties of their distance estimates,
NGC 6723 and NGC 6522 might be positioned on either the far side or 
near side of the GC, in which case the velocity components
$\Pi$ and $\Theta$ change signs.
Thus retrograde orbits become prograde and vice versa; 
however this does not alter
the large inclination of their orbits. In other words, the sense of rotation
with respect to the disk is of little significance
when orbits are nearly polar (Tab. 5).

NGC 6266 has a large rotation velocity, while the other two components are
relatively low (Tab. 4). This velocity vector, combined with
a galactocentric distance $R_{GC} = 1.4$ kpc (Tab. 4) indicate that
NGC 6266 belongs to a rotationally supported system.
Its rotation velocity is somewhat larger than that derived for a solid body
rotating with the Galactic bar's pattern speed (Tab. 4).
As the notion of a kinematical stellar disk is inappropriate at
these galactocentric distances, we can think of NGC 6266
as a cluster that belonged to a disk system before the bar formed,
rather than to a pressure-supported system such as the halo.
In this sense, NGC 6266 reminds us
of thick-disk, metal-poor clusters such as 
NGC 6626, NGC 6752, and perhaps NGC 6254 (see Paper III).
These clusters represent the metal-weak thick disk 
(MWTD), found to be
very prominent in the field stellar component within 1 kpc of the Sun
(Beers {\it et al.} 2002 find a fraction of 30-40\% MWTD at
metallicities $<$ -1.0). The novelty provided by the clusters is
that this component can be probed at smaller galactocentric
 radii than local samples.

\subsection{Metal Rich ([Fe/H] $ > -1.0$) Clusters}

Based on spatial distribution, radial velocities and masses, 
BS97 have proposed 
a new classification of clusters with  
[Fe/H] $ > -0.8$.
The high-mass system  (M$_V < -7.9$) is thought to be
representative of the bulge. It has an angular 
velocity $\omega = 15$ km s$^{-1} $ kpc$^{-1}$, and 
a line-of-sight velocity dispersion  $\sigma_{los} = 56$
 km s$^{-1}$, assuming  a solid-body rotation in a
Frenk \& White (1980) type solution. 
The low-mass system (M$_V \ge -7.9$) is a sum of two components: 
a bar-like component, and a thick-disk-like component.
The bar-like  component has an angular velocity ($\sim 59$ 
km s$^{-1} $ kpc$^{-1}$) similar to traditional 
values of Milky Way bar's pattern speed, and a  $\sigma_{los} = 55$
 km s$^{-1}$.
The thick-disk-like component resides at $R_{GC} \sim 5$ kpc, and 
it has typical thick-disk kinematics (see Table 1 in BS97).

According to BS97, by mass and metallicity,
NGC 6316 is a bulge cluster (Table 5).
If we use the angular velocity of the bulge globular-cluster system 
defined in BS97
to predict the rotation velocity at NGC 6316's location, we 
obtain that NGC 6316's retrograde velocity is at 2.2$\sigma$
of that predicted by the solid body.
To estimate $\sigma$ we have used the uncertainty in the
$\Theta$ component (Tab. 4) and we have taken the 
velocity dispersion along $\Theta$ to be equal to
the line-of-sight velocity dispersion derived from the
solid-body solution for all of the candidate bar clusters in BS97.
The constant-rotation velocity solution in BS97  places NGC 6316
within 1.6$\sigma$ of the mean of the globular-cluster
bulge system defined by BS97, as the line-of-sight velocity dispersion is
$\sim 70$  km s$^{-1}$.
A similar comparison with the BS97 bar system, places NGC 6316 
at 4.3$\sigma$ from the rotation velocity expected from the solid body.
Thus, according to its velocity,
NGC 6316 can be thought of
as a bulge cluster rather than a bar cluster, and this is in agreement with 
its classification according to metallicity and mass.

Both NGC 6304 and NGC 6553 have large rotation velocities and low to moderate
radial and vertical velocity components (Tab. 4).
The $\Theta$ components are somewhat larger than the rotation velocities 
expected from a solid body that has an angular velocity 
equal to the Galactic bar's pattern speed (Tab. 4).
The large $\Theta/V'$-velocity component of NGC 6553 may be due to the
assumption that the reference stars (bulge stars) in the 
proper-motion solution have zero velocity (see Section 3.1).
The Galactic longitude of this cluster is larger than that of NGC 6522, and
6528, therefore this assumption may not be correct. In fact, at
a longitude of $ \sim 5\arcdeg$ (Table 3), 
the bulge stars are expected to have 
a rotation of 45 km s$^{-1}$ (Tiede \& Terndrup 1999 measure
a projected mean rotation of 9 km s$^{-1}$ deg$^{-1}$). Thus, the 
rotation-velocity component should, more likely, be $V' \sim 180$ km s$^{-1}$,
and correspondingly, $\Theta = 175$  km s$^{-1}$, a value that is 
much closer to NGC 6304's rotation velocity.
Masses and metallicities assign these clusters to the bar system of
BS97 (NGC 6553 is slightly more massive than the BS97 limit; see Tab. 5).
Their kinematics, at face value, are not inconsistent with 
bar membership, especially if we must abandon the idea of
disk-like orbits inside the corotation radius. These clusters could 
very well be members of the former disk/thick-disk 
transformed into a bar in the
inner regions. They can be thought of as the extension of
the 5-kpc thick disk clusters in BS97 into the inner Galactic
regions. What is intriguing is that NGC 6266, a massive, metal-poor cluster,
is kinematically very similar to NGC 6304 and NGC 6553 (Tab. 4, 5;
Fig. 4). The slight difference is that NGC 6266, while 
currently located farther from the Galactic plane than NGC 6304 and
NGC 6553 (Fig. 4), moves away from the plane
with the largest velocity of the three clusters (Tab. 4).
Strictly speaking, this implies that NGC 6304 and 6553 are
in a more flattened system than NGC 6266.

NGC 6528 qualifies as a bar cluster according to BS97 (Tab. 5).
Its $\Theta$ velocity component is in very good agreement with
the rotation of a solid body that has the Galactic bar's pattern speed
(Tab. 4). 
The very low vertical velocity together with
the close distance to the Galactic plane (Fig. 4) indicate that the
cluster will spend its time near the Galactic plane, while the large
$U$-velocity component, indicates a radial orbit.
We note that NGC 6528's distance has a wide range of values, 
even from more recent determinations 
(e.g., Feltzing \& Johnson 2002 determine a distance of 7.2 kpc, while
HR99 give a distance of 9.3-10.7 kpc from isochrone 
fitting, and from the horizontal-branch metallicity relation, respectively).
Even if, according to distance errors, NGC 6528 is displaced from the
far side of the bulge to the near side, the $U$-velocity component will remain
practically the same (while the $\Pi$ component changes sign,
but $\Theta$ does not,  see Fig. 4).
This will leave the character of the orbit unchanged.

These kinematical arguments together with the fact that
NGC 6528 is, arguably, the most metal rich Galactic globular cluster
(Feltzing \& Johnson 2002),
make a strong case for a globular formed within the bar.

We note that, the two clusters that are located outside the bar 
limits, NGC 6316 and 6723 (Fig. 4),
also have kinematics inconsistent with that of the bar. 
Among the clusters located within the bar limits,
only NGC 6522 --- a metal poor cluster ---
seems to have kinematics inconsistent with 
that of the bar.

\section{Summary}

We have measured absolute proper motions on the 
$Hipparcos$ system for four globular 
clusters located in the bulge region.
In addition to these, we include in our discussion three clusters
located in the bulge region that had previous proper-motion measurements with
respect to bulge stars. We caution that velocities determined 
from these latter proper motions can be regarded as with respect 
to an inertial frame, only under strict assumptions for the location of the
cluster.

Without a detailed orbital analysis that we defer for a future
paper, we discuss the kinematics of these clusters in relation with
their metallicities and masses. All of the clusters have velocities
smaller than the escape velocity from the bulge, indicating that
they have orbits that never leave the bulge region.
From the three metal-poor
clusters in our sample, two of them, namely NGC 6552, and NGC 6723
have kinematics consistent with halo membership. The third, NGC 6266,
seems to belong to a rotationally supported system 
(reminding one of the thick disk, but not necessarily thick disk at these
GC distances)
rather than to a pressure supported one. 
Of the four metal rich clusters, two clusters 
NGC 6304, and NGC 6553 have kinematics resembling a rotationally
supported system, much like NGC 6266. NGC 6528's velocity indicates a 
radial orbit, confined to the Galactic plane.  It's rotation velocity is 
consistent with the velocity predicted from the rotation of a solid body
that has the angular pattern speed of the Milky Way bar. 
As a low-mass, and arguably the most metal-rich Milky Way cluster, 
NGC 6528 may be a genuine bar cluster. The fourth metal rich cluster,
NGC 6316 has kinematics consistent with membership to a hotter 
system than the bar.

This work is supported in part by NSF grant AST-0098687.

\newpage

\begin{figure}
\caption{The correction to absolute proper motion
for plate-pair solutions 506Y1Y2 and 506y1-ucac,
left and right panel respectively. Labels represent the various
image orders as described in the text. The closed symbol is
the weighted mean of the measurements.}
\end{figure}

\begin{figure}
\caption{Relative proper motions of stars in the region of
NGC 6304 as a function of magnitude, for visual-plate solutions
in SPM fields 506 and 507. Horizontal
 lines represent the (centering-error) weighted 
means.}
\end{figure}

\begin{figure}
\caption{Absolute proper-motion results for each cluster.
Open symbols show the various plate-pair solutions, and the
closed symbols represent the final adopted absolute
proper motion.}
\end{figure}

\begin{figure}
\caption{Projection on to the Galactic plane (top)
and perpendicular to the Galactic plane (bottom) 
of the positions and velocities of the clusters.
Open symbols are the metal poor clusters, while closed symbols are the 
metal rich clusters. The ellipse represents the Galactic bar
with geometric parameters taken from Bissantz \& Gerhard (2002).
The GC is represented with a cross, and the dotted circle in the top panel
marks a radius of 2.7 kpc, as an indication of the bulge region.}
\end{figure}

\clearpage

{\sc TABLE} 1. Plate-pair Solutions

{\sc TABLE} 2. Absolute Proper Motion Results

{\sc TABLE} 3. Cluster Data

{\sc TABLE} 4. Velocities

{\sc TABLE} 5. Cluster Parameters

\end{document}